# Holon-Doublon Binding as the Mechanism for the Mott transition

Peter Prelovšek[1,2], Jure Kokalj[1], Zala Lenarčič[1], and Ross H. McKenzie[3]

[1]*J. Stefan Institute, SI-1000 Ljubljana, Slovenia*

[2]*Faculty of Mathematics and Physics, University of Ljubljana, SI-1000 Ljubljana, Slovenia and*

[3]*School of Mathematics and Physics, University of Queensland, Brisbane, 4072 Queensland, Australia*

We study the binding of a holon to a doublon in a half-filled Hubbard model as the mechanism of the zero-temperature metal-insulator transition. In a spin polarized system a single holon-doublon (HD) pair exhibits a binding transition on a 3D lattice, or a sharp crossover on a 2D lattice, corresponding well to the standard Mott transition in unpolarized systems. We extend the HD-pair study towards non-polarized systems by considering more general spin background and by treating the finite HD density within a BCS-type approximation. Both approaches lead to a discontinuous transition away from the fully polarized system and give density correlations consistent with numerical results on a triangular lattice.

PACS numbers: 71.27.+a, 71.30.+h, 71.10.Fd

## I. INTRODUCTION

The mechanism of the Mott metal-insulator transition (MIT)[1], i.e., the transition in a half-filled band induced by the electron-electron repulsion, has been a major theoretical and experimental challenge for the last few decades[2]. The primary signature of the MIT (at zero temperature $T=0$) is the appearance of the charge gap $\Delta_c$ and vanishing of the charge susceptibility $\chi_c$ on the insulating side. A number of distinct mechanisms for the transition have been proposed. These include those due to Brinkman and Rice[3] (where the quasi-particle weight in the metallic phases approaches zero as the transition is approached), Hubbard (where vanishing of the charge gap occurs when the upper and lower Hubbard bands overlap), or Dynamical Mean-Field Theory (DMFT)[4] which combines both these features. Powerful numerical approaches have been applied to prototype models such as the Hubbard model, in particular methods emerging from DMFT and variational Monte Carlo approaches[5]. The physics of the MIT is complicated and challenging due to the possibility of coexisting antiferromagnetic (AFM) spin orderings within the Mott insulator, depending on the underlying lattice. The effect of the competing spin order on the MIT is expected to be less pronounced in non-bipartite lattices, as in the Hubbard model on a triangular lattice[5–9], of relevance to organic charge-transfer salts[10–13]. Recently also fullerenes[14,15] have become of interest as an example of the MIT in 3D frustrated lattices.

In this paper we show that within the half-filled single-band Hubbard model the MIT can be understood as a transition from an unbound to a bound state of doubly occupied sites (doublons) and empty sites (holons). This view has been advocated long ago[1,6,16–18], but made more explicit in more recent variational[5,19], field theory[20], and slave-boson[21] approaches. The concept sketched for a triangular lattice in the right panel of Fig. 1 is simple: in a half-filled band one can consider the holon-doublon (HD) pairs as basic charge excitations relative to the reference state with singly occupied sites. Within the Mott insulator at large Coulomb repulsion $U > U_c$ the HD pairs (at $T=0$) are bound while in a metal at $U < U_c$ pairs are not bound and holons and doublons separately contribute to electrical current and to the finite charge stiffness $D_c > 0$. Such an approach to the MIT becomes even more reasonable if one considers a polarized half-filled band, i.e. the system with nearly saturated spin magnetization $m \sim 1/2$ due to, e.g., the presence of a uniform magnetic field. In this case one deals with a simplified problem of vanishing (but non-zero) concentration of HD pairs[22], which can be treated exactly on any lattice model. At half-filling introducing the magnetization ($|m| < 1/2$) as an additional degree of freedom opens the possibility to explore the ground-state phase diagram of the Hubbard model for any particular lattice.

In particular, this approach allows for a novel path to the challenging regime of $m \sim 0$, as well as a benchmark and an alternative interpretation of the MIT. In Fig. 1 we anticipate the result for the Hubbard model on a triangular lattice, which will be our focus later on. Extending the study of the MIT to Hubbard systems in large magnetic fields[23,24] is of experimental relevance, e.g., for organic conductors in a magnetic field[25] and for cold fermions in optical lattices. We should mention here the relation to several studies of the attractive Hubbard model[26–29], which can be mapped to the repulsive model in a magnetic field[30], provided that the lattice is bipartite.

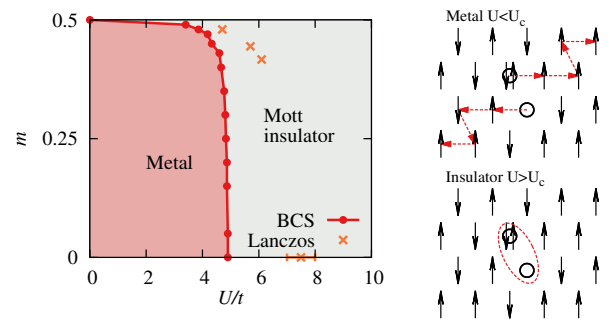

Figure 1. (Color online) Phase diagram of the Hubbard model on a triangular lattice at half filling as calculated from the BCS-type approximation and exact diagonalization (Lanczos) results. The Lanczos results are for magnetization $m \neq 0$ calculated at $N_d$ holon-doublon pairs and system sizes $N$ with values $N_d/N = 2/100, 2/36$ and $3/36$, while result for $m = 0$ is from Ref. 8. Right panels show a schematic unbound (bound) holon-doublon pair in the metallic (insulating) phase on a triangular lattice with a non-zero spin polarization.

The paper is organized as follows: In Sec. II we employ an analysis of the single HD pair in a spin polarized background $m \sim 1/2$. Concentrating on non-bipartite lattices, in particular the triangular lattice, this simplified case gives the critical (or crossover) $U_c$ for the MIT transition and also the dependence of the charge gap $\Delta_c(U)$, and double occupancy $\tilde{D}$ that are quite close to the numerical results for the unpolarized system. The analysis is extended towards the non-polarized system in two ways. On one hand we consider in Sec. III the motion of a HD pair within a more general spin background, for which the MIT is a discontinuous one. In an alternative approach we deal in Sec. IV directly with a finite density of HD pairs within a BCS-type approximation[6,26,31], giving a fair agreement with numerical finite-system results. Conclusions and implications are presented in Sec. V.

## II. SINGLE HOLON-DOUBLON PAIR BINDING

We consider the single-band Hubbard model with nearest-neighbor (n.n.) hopping on a general lattice,

$$H = -t \sum_{\langle ij \rangle \sigma} (c^\dagger_{j\sigma} c_{i\sigma} + c^\dagger_{i\sigma} c_{j\sigma}) + U \sum_i n_{i\uparrow} n_{i\downarrow}. \quad (1)$$

We restrict our study to a half-filled band and $T = 0$. To analyze the MIT we consider the system with a finite magnetization $m = S^z_{tot}/N$, where $S^z_{tot} = \sum_i S^z_i$ is the total spin, and $N$ the number of lattice sites. Clearly, the case with maximum $m = 1/2$ is a ferromagnetic (FM) band insulator, but the problem is nontrivial for any $|m| < 1/2$.

We first consider the problem within the sector $S^z_{tot} = N/2 - 1$, i.e. a single HD pair. (we note that a similar case has been considered also by Kohn[16]). The problem within the model (1) is that of two particles with total momentum $q$,

$$|\psi^r_q\rangle = N^{-1/2} \sum_l e^{iql} |\varphi_{l,l+r}\rangle, \quad |\varphi_{lm}\rangle = c^\dagger_{l\downarrow} c_{m\uparrow} |FM\rangle, \quad (2)$$

and the onsite attraction $-U$, leading to the gap equation for the pair energy $E_q$[22],

$$1 = -\frac{U}{N} \sum_k \frac{1}{E_q - U - \eta_k}, \quad \eta_k = \epsilon_{q-k} - \epsilon_k. \quad (3)$$

Here $\epsilon_k$ is the single-electron band energy. Eq. (3) yields the energy $E_q$ of the bound HD pair at given $q$ provided that $E_q < U + \min_k[\eta_k]$. In the opposite case, the pair is unbound and states form a continuum. For a general lattice the minimum for the bound HD pair is $E_0 = \min[E_q] = E_{q_0}$ whereby $q_0$ is straightforwardly determined for standard cases.

### A. Hypercubic lattices

In a $D$-dimensional hypercubic lattice one gets, due to the particle-hole symmetry, $q_0 = \pi(1, \ldots, 1)$ and $\epsilon_{q_0-k} = -\epsilon_k$ so that $\eta_k = -2\epsilon_k$. The continuum of states has the lower edge at $E_1 = U - 2zt$, $z$ being the number of n.n., so that the charge gap

$$\Delta_c = E_1 - E_0 = U - 2zt - E_0. \quad (4)$$

In the insulating regime ($U \gg t$) it follows that $E_0 \sim -4zt^2/U$, and so $\Delta_c \sim U - 2zt$[1]. Finite $U_c \sim 2zt > 0$ appears in a 3D case. On the other hand, in 2D there is strictly no transition due to the singular (step-like) single-electron density of states (DOS) $\mathcal{D}(\epsilon)$ at the band edge. The latter in connection with Eq. (3) generally leads for 2D to $U_c = 0$. Still there is a sharp crossover in $\Delta_c(U \sim U^*_c)$ with $\Delta_c$ being exponentially small for $U \lesssim U^*_c$. The same effect is present also within the triangular lattice discussed further-on with the result for $\Delta_c$ presented in Fig. 2.

The $m \sim 1/2$ case should be distinguished from $m = 0$, where, e.g., on a square lattice $U_c = 0$ due to Fermi surface nesting[32,33]. Here, we should mention a related limit for the $z \to \infty$ Bethe lattice, relevant for DMFT studies, where the central input is the DOS

$$\mathcal{D}(\epsilon) = (2/\pi W) \sqrt{W^2 - \epsilon^2}, \quad (5)$$

with $W$ being the effective half-bandwidth. In this case one finds analytically that $U_c = W$, serving as a nontrivial benchmark for numerical studies[29,34].

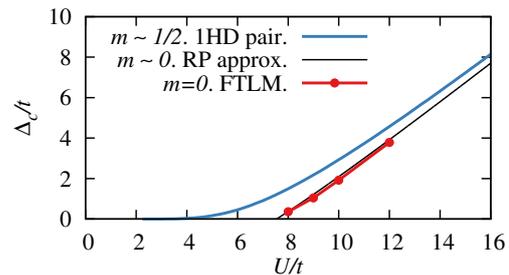

Figure 2. (Color online) Charge gap $\Delta_c$ vs. $U$ for the Hubbard model on triangular lattice. The results shown are calculated for a single holon-doublon (HD) pair at $m \sim 1/2$ (thick blue line), a single HD pair in an unpolarized spin background $m = 0$ within the retraceable-path approximation (thin black line), and numerically for $m = 0$ (red line with points).

### B. Triangular lattice

Let us now turn to the non-bipartite lattices, where a widely studied case is the Hubbard model on a triangular lattice with n.n. hopping. $E_q$ reaches a minimum value at $q_0 = (4\pi/3, 0)$. Since we are dealing with 2D DOS there is strictly no MIT but rather a sharp crossover at $U^*_c \sim 6t$, well visible in $\Delta_c(U)$, presented in Fig. 2. We can judge the relevance of this exact result for $m \sim 1/2$ by comparing it with numerical results for the case $m = 0$, obtained by using the finite-temperature Lanczos method (FTLM) on $N = 16$ sites (taken from Fig. 2 in Ref. 8 and multiplied by 2 due to different definitions of the gap). In spite of quite different spin background, Fig. 2 shows a fair correspondence between the single-HD result (at least for $U/t > 10$) and $\Delta_c(U)$ at $m = 0$. Clearly, one should take into account that at $m \sim 1/2$ the gap $\Delta_c(U)$ exhibits only a crossover at $U_c^*$ (with exponentially small gap for $U < U_c^*$) while strictly $U_c = 0$ for $m = 1/2$. One should bear in mind that also other numerical results for triangular lattice show considerable uncertainties, see e.g. the compilation in the Supplementary material for Ref. [8].

For a single HD pair it is straightforward to evaluate also the actual doublon density $\tilde{D}$, defined by the probability that the holon and doublon are not on the same site, leading to the explicit expression,

$$\tilde{D} = 1 - \Big(\frac{1}{N}\sum_k \frac{U^2}{(E_0 - U - \eta_k)^2}\Big)^{-1}. \quad (6)$$

We calculate double occupancy as $D = \tilde{D}/2$ to have correct normalization per site in order to compare with $D$ standard for $m = 0$. We then compare it with other results for $m < 1/2$ in Fig. 3.

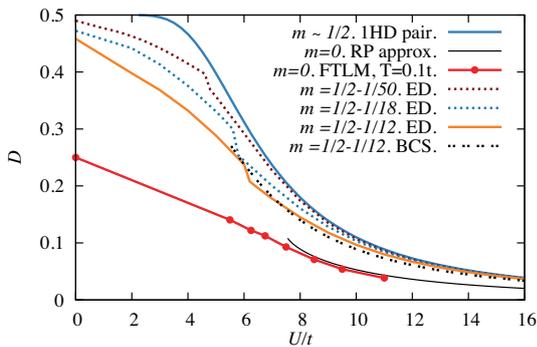

Figure 3. (Color online) Double occupancy $D$ (per double number of pairs $2N_d$) vs. $U$. The results shown are calculated for $m \sim 1/2$ with a single HD pair (denoted 1HD), for a single HD pair within the RPa (denoted RP approx.), with FTLM and taken from Ref. 8 (denoted FTLM), with exact diagonalization at $T = 0$ and various $m$ (denoted ED), and with BCS approximation for particular $m$ (denoted BCS).

### III. HOLON-DOUBLON PAIR IN A GENERAL SPIN BACKGROUND

An extension from $m \sim 1/2$ towards $m = 0$ can be made by keeping a single HD pair, but formed in a more general spin background. The reference state should be first found within the spin Heisenberg model describing the Mott insulating phase. Being primarily interested in frustrated lattices we assume that spin correlations are of short range. In such a case the retraceable path approximation (RPa) is appropriate[35], excluding the HD motion along loops. Within the RPa one describes a HD pair created/annihilated on the same site with the local Green's function,

$$G(\omega) = \frac{1}{\omega - \Sigma(\omega)}, \qquad \Sigma(\omega) = 2\alpha z t^2 G_2(\omega), \quad (7)$$

where the recombination hopping is renormalized by the spin correlation factor

$$\alpha \equiv \langle\Phi_0|(\frac{1}{2} - 2\mathbf{S}_i \cdot \mathbf{S}_j)|\Phi_0\rangle = \frac{1}{2} - 2\mu. \quad (8)$$

Here $i, j$ denote nearest neighbors. Further hoppings of either holon and doublon are taken into account with $G_2(\omega)$. Assuming that both holon and doublon hop independently along the retraceable paths and neglecting spin string energy, $G_2(\omega)$ is[35],

$$G_2(\omega) = \int\int d\epsilon_1 d\epsilon_2 \frac{\mathcal{D}(\epsilon_1)\mathcal{D}(\epsilon_2)}{\omega - U - \epsilon_1 - \epsilon_2}, \quad (9)$$

where the DOS within the RPa is given for the corresponding Bethe lattice, Eq. 5, with $W = 2\sqrt{z-1}t$.[36] The bound solution ($E_0 = \Sigma(E_0)$) is stable provided that it is below the continuum of unbound HD-pair states at $E_1 = U - 2W$ derivable from Eq. (9). The corresponding energy difference $\Delta_c = E_1 - E_0$ is interpreted as the charge gap $\Delta_c$ relevant for the MIT. In contrast to the single HD pair in the FM background, Eq. (3), the MIT is a discontinuous one due to energy level crossing of bound and unbound HD-pair solutions. It should be mentioned that direct comparison with the $m \sim 1/2$ is not feasible since loop trajectories would be required to represent fully the latter case. In Fig. 2 we plot the resulting $\Delta_c(U)$ taking $\mu = -0.182$ for $\alpha$ as appropriate for the Heisenberg model that holds for $U \gg t$[37–40]. It is evident that the RPa result is closer to the numerical one for $m = 0$, which is obtained by FTLM and shows small finite size effects[8]. This agreement gives support to the RPa for triangular lattice. Within the same method it is straightforward to evaluate also the effective double occupancy (per HD-pair) $\tilde{D}$ evaluated from Eq. (7) via the weight of bound-state pole

$$\tilde{D} = 1 - \Big[1 - \frac{\partial\Sigma}{\partial\omega}|_{\omega=E_0}\Big]^{-1}, \quad (10)$$

or from $\partial E_0/\partial U$. We present the results in Fig. 3, where it can be seen to compare favourably with the FTLM result for $m = 0$.

### IV. FINITE HOLON-DOUBLON PAIR DENSITY

An alternative approach to the case of a finite HD pair density $N_d/N$ with maximal number of doublons $N_d$ and holons $N_h = N_d$, and $m = 1/2 - N_d/N$, is to start from a reference FM state (all spins up) and to define doublon $d_i^\dagger = c_{i\downarrow}^\dagger$ and holon $h_i^\dagger = c_{i\uparrow}$ operators, and rewrite the repulsive Hubbard model, Eq. (1), as an HD attractive one[6,30,41],

$$H = \sum_k (\epsilon_k^d d_k^\dagger d_k + \epsilon_k^h h_k^\dagger h_k) + UN_d - U\sum_i n_{di} n_{hi}. \quad (11)$$

Here, $\epsilon_k^d = -\epsilon_k^h = \epsilon_k$. This should be distinguished from the standard negative-$U$ model with fermions having the same dispersion. Due to attraction between holons and doublons it is natural to approximate the ground state with a mean-field BCS-type approach, since it captures HD pair binding and their coherent superposition.

### A. BCS-type approach

As in several previous works[6,26,31,42] we approximate the ground state of the collective HD pair state with a BCS-type wavefunction, representing the coherent superposition of bound HD pairs,

$$|\Psi_0\rangle = \prod_k (u_k + v_k b_k^\dagger)|FM\rangle \tag{12}$$

where $b_k^\dagger = d_k^\dagger h_{q-k}^\dagger$ is the HD pair creation for given $q$ and

$$(u_k^2, v_k^2) = (1 \pm \zeta_k/E_k)/2. \tag{13}$$

Here, $|FM\rangle$ is the reference FM state of all spins up as already used in Eq. (2). $\zeta_k = \eta_k/2 - \bar{\mu}$ (with $\eta_k$ from Eq. (3)) is the HD pair dispersion, $\bar{\mu} = (\mu_d + \mu_h)/2$ is the effective pair chemical potential, $E_k = \sqrt{\zeta_k^2 + \Delta^2}$ the quasi-particle excitation energy and $\Delta$ the pairing order parameter. The BCS ground state energy is then given by

$$E_{BCS} = \sum_k (\zeta_k - E_k) + \frac{N}{U}|\Delta|^2 + [U(1-\bar{n}_d) + 2\bar{\mu}]N_d,, \tag{14}$$

where we used the abbreviation $\bar{n}_d = N_d/N$. We note that $|\Psi_0\rangle, u_k, v_k, E_k, \zeta_k$ and $\Delta$ depend also on $q$, and we omit this from our notation for clarity. At fixed $\bar{n}_d$ (or $m = 1/2 - \bar{n}_d$) one still has to solve self-consistently equations for $\bar{\mu}$ and $\Delta$,

$$\bar{n}_d = \frac{1}{N}\sum_k v_k^2, \qquad 1 = \frac{U}{2N}\sum_k \frac{1}{\sqrt{\zeta_k^2 + \Delta^2}}. \tag{15}$$

In addition, $q$ that gives the lowest energy $E_{BCS}$ has to be determined. Nevertheless, we consider here only $q = q_0 = (4\pi/3, 0)$, which is clearly the solution for $m \sim 1/2$ and also away from $m = 0$[6].

$\Delta$ does not have the meaning of the charge gap[26] like in the usual weak-coupling BCS theory for superconductors. For the MIT one needs to compare $E_{BCS}$ at a given $\bar{n}_d$ with the metallic state or HD liquid. The energy of the metallic state $E_M$ is obtained from Eq. (14) by putting $\Delta = 0$ and filling the doublon and holes states up to $\mu_d$ and $\mu_h$, respectively.

The MIT appears as a crossing of $E_{BCS}$ and $E_M$ at $U = U_c$ or the appearance of positive condensation energy $E_c = E_M - E_{BCS}$ with increasing $U$. The transition is of first order due to jump of $\Delta$ at this crossing. Within BCS one can calculate also the double occupancy $D$ (normalized to $2N_d$) as

$$D = \frac{1 - \bar{n}_d}{2} - \frac{\Delta^2}{2\bar{n}_d U^2}. \tag{16}$$

It is important to realize that within a BCS approximation one can reproduce the exact single HD pair results. We note that within the dilute limit, $\bar{n}_d \ll 1$, where also $\zeta_k \gg \Delta$, Eqs. (15) reduce to the form for a single HD pair, Eq. (3), on any lattice. The relation is derived by identifying $\bar{\mu}$ with the single HD pair energy $E = U + 2\bar{\mu}$ obtained within the $\bar{n}_d \ll 1$ limit of Eq. (14). In this case, it also follows

$$\zeta_k \sim U/2, \qquad \Delta \sim 2\zeta_k\sqrt{\bar{n}_d} \sim U\sqrt{\bar{n}_d} \ll t, \tag{17}$$

in analogy to bipartite lattices[26,30].

### B. Comparison with numerical results

Within BCS one can also calculate the double occupancy $D$, Eq. (16), and we compare the result in Fig. 3 with the exact diagonalization (ED) result obtained with the Lanczos method[43] on a finite cluster for $N_d/N = 1/12$ ($N_d = 3$, $N = 36$) and find good agreement. ED results are shown also for $N_d = 2$ HD pairs on $N = 36$ and $N = 100$ sites, respectively.

The validity of the BCS approximation at low HD densities $N_d/N \ll 1/2$ can be tested also via density correlation functions. Of particular interest are the doublon-holon and doublon-doublon (or holon-holon) density correlations

$$C_{dh}(r) = \sum_i \langle \tilde{n}_{di}\tilde{n}_{h,i+r}\rangle, \quad C_{dd}(r) = \sum_i \langle \tilde{n}_{di}\tilde{n}_{d,i+r}\rangle, \tag{18}$$

where $\tilde{n}_{di} = n_{di}(1-n_{hi}), \tilde{n}_{hi} = n_{hi}(1-n_{di})$ are 'true' (projected) doublon and holon operators, respectively. We present the comparison between results from ED at $N_d = 2$ and the BCS function also projected onto $N_d = 2$ HD pairs. Fig. 4 demonstrates the effective HD attraction via $C_{dh}$ and the repulsion between doublons via $C_{dd}$. The agreement between ED and the BCS approximation is remarkable within the insulator regime $U/t = 6, 8 > U_c/t$. Regarding correlations, the main signature of the insulating (HD binding) regime is that $C_{dh}$ and $C_{dd}$ converge (according to the BCS approximation exponentially) at larger $r$. In contrast, in the metallic case the approach is expected to be much slower (algebraic) and indeed for $U = 4t$ the $C_{dh}$ remains larger than $C_{dd}$ even for the largest $r$ in a given system indicating that the holon is not bound to the doublon.

Finally let us discuss the phase diagram of the Hubbard model on a triangular lattice, presented already in Fig. 1. In the limit $m \sim 1/2$ the BCS insulating state is stable for any $U > 0$. On the other hand, for any $m < 1/2$ the condensation energy $E_c(U)$ is vanishing at finite $U = U_c$. For comparison we present in Fig. 1 also ED results, where the MIT transition is monitored by the discontinuity of the double occupancy $\tilde{D}(U)$. Value of $U_c(m = 0)$ is taken from FTLM result presented in Ref. 8.

## V. CONCLUSIONS AND IMPLICATIONS

The aim of this work is to present a theory of the Mott MIT within a half-filled Hubbard band starting from a spin polar-ized system. The advantage of such an approach is that one can systematically follow the MIT by reducing the polarization by which the behavior evolves from the exactly solvable (but nontrivial) single HD-pair binding at $m \sim 1/2$ into a collective HD-pair phenomenon going towards $m \sim 0$. Our consideration of $m < 1/2$ requires some approximations (RPa and BCS) but we show that they produce meaningful and quite reliable results also for $m \sim 0$. Within these approximations the MIT at $m < 1/2$ is discontinuous, and the following simple scenario is suggested by the BCS. When $U$ is reduced from large values where the density and size of HD pairs are small, the pairs start to increase in size and the kinetic energy decreases. When the pairs start to overlap and due to the hard core repulsion between them (see $C_{dd}$ in Fig. 4) the kinetic energy cannot be further reduced, the metallic state becomes preferred, leading to a discontinuous MIT.

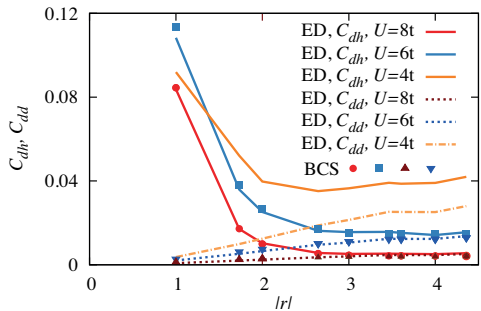

Figure 4. (Color online) Doublon-holon $C_{dh}$ and doublon-doublon $C_{dd}$ correlations vs. distance $r$ for different $U/t$ evaluated for two HD pairs, $N_d = 2$, on $N = 64$ sites. Results obtained by using the full ED (lines) can be compared to the ones evaluated within the BCS wavefunction for $U/t = 6, 8$ (points).

Concentrating on a triangular lattice we have shown that the charge gap $\Delta_c(U)$ for a single-HD pair is quite similar to that for $m \sim 0$, bearing in mind that $\Delta_c(U)$ at $m \sim 1/2$ appears only as a crossover. Similar agreement is seen for the (properly normalised) double occupancy $D(U)$. When we consider the agreement with the numerical results, we should point out that a part of the uncertainties are due to the approximations (RPa and BCS) used. Moreover, numerical benchmarks for the MIT (available in the literature primarily for $m \sim 0$) in the triangular lattice are even more scattered (see the compilation in Ref. 8), so it is at this stage hard to better quantify our HD-pairing results.

On the other hand, it is evident that presented RPa and BCS approaches have some restrictions when considering unpolarized $m \sim 0$ regime. Both approximations deal primarily with the bound pairs, i.e. the insulating phase, while the metallic side of the MIT is considered only on the simplest level, which can to our opinion underestimate the critical $U_c$ of the MIT at $m \sim 0$. Within RPa we did not take into account the influence and competition of different spin orders close to the MIT, which should be less severe for frustrated lattices. With increasing concentration of HD pairs the interplay of close pairs is not taken into account within the RPa approach. However, it should be stressed that the criterion for the latter inter-pair effects is not directly $N_d/N$, but rather $D$ which measures the density of 'real' doublons and holons. In this respect, we notice that at $m \sim 0$ $D \sim 0.1 \ll N_d/N$ is small even at the MIT, as evident for the triangular lattice, e.g. in Fig. 3.

Concerning the BCS approximation it has two essential ingredients, important to describe the holon-doublon (HD) insulator at finite HD density $\bar{n}_d$: a) BCS wavefunction represents the coherent superposition of bound HD pairs, b) it takes into account the Pauli exclusion principle at arbitrary $\bar{n}_d$. The latter clearly leads to proper (repulsive - like) fermionic correlations $C_{dd}$ and $C_{hh}$. The correlations $C_{hd}$ are also taken on the two-body level. Still, it is evident that all correlation effects are not taken into account within the BCS wavefunction. The BCS approximation is also better justified for more extended pairs, while its non-straightforward limitations are similar to the strong-coupling BCS (see Ref. 26). BCS approximation also neglects quantum fluctuations of the gap order parameter, which could destabilise the insulating phase in some regime. It in addition yields the condensation of HD pairs, which is not necessary for MIT. It therefore remains to be checked to what extent are the BCS results, as e.g. $C_{hd}$, quantitatively reproduced for $\bar{n}_d \gg 0$. This is, however, a challenging task also for the present status of numerical methods. The validity of the BCS phase diagram in Fig. 1 evidently depends on the quality of the BCS wavefunction, but also on the assumed simple form of the reference metallic state energy $E_M$. Still, the sharp saturation of $U_c(m \sim 0.5)$ has (at least partly) the same origin to 2D singularity DOS, leading also to the very sharp crossover stemming from Eq.(3) and showing up in the $\Delta_c(U)$ in Fig. 2. While further nearly constant $U_c(m)$ emerging from BCS equations (calculated at fixed $q = q_0$) has no evident explanation, it appears that only gross features of the band matter, as is also the case e.g. for $\Delta_c(U > U_c^*)$ in Fig. 2.

In our study we concentrated on the charge sector, being the driving mechanism of the MIT. It is however evident that away from $m \sim 1/2$ in the insulating phase the spin long-range order, e.g. the spiral $\langle \mathbf{S}_q \rangle$ or an AFM order, can emerge. This appears to be the case also for the extensively studied frustrated triangular lattice, at least for $U \gg t$[7,37,38,44–51]. While it is evident that for bipartite lattices the nesting mechanism leads to the AFM order at $m = 0$ for any $U > 0$, the question of possible spin order in frustrated lattices remains challenging, even for numerical methods. Still we should point out that the spin order is a secondary phenomenon for the MIT since it involves only exchange energies of the order of $J = 4t^2/U$, whereas the charge sector leads to the MIT due to the HD attraction $U$. We note that HD binding is also prerequisite for the description with Heisenberg type models, which are typical starting points for the discussions of spin ordering.

The character of the MIT transition at $m \sim 1/2$ depends on the lattice dimensionality. While in the 2D lattice the single HD-pair exhibits only a rather sharp crossover, the binding transition becomes well defined and continuous in 3D lattices. The most intriguing is the f.c.c. lattice with $U_c \sim 12t$ being also of experimental relevance to fullerides[15] and a natural extension of the presented work. Finally, we comment on the relation to other scenarios (or interpretations) of the Mott MIT. We are close to the original Mott proposal[1] for the charge gap as $\Delta_c = U - 2zt$. Similar closing of the gap between Hubbard bands is incorporated also in the DMFT theories. Still, the DMFT approach, as well as the Brinkman-Rice picture, is based on single-particle properties (the one-electron Green's function) and the relation to the HD binding, a two-particle phenomenon is less evident. In any case, the relation of our HD-pair binding approach to existing theories of the MIT should be further explored.

## ACKNOWLEDGMENTS

The authors acknowledge fruitful discussions with T. Tohyama, R. Žitko, J. Mravlje and P. Phillips. This work was supported by the Program P1-0044 and projects J1-4244 and Z1-5442 of the Slovenian Research Agency (ARRS) and a Discovery Project grant from the Australian Research Council. Z.L. is supported also by the L'Oréal-UNESCO national scholarship "For Women in Science".

[1] N. F. Mott, Proc. Phys. Soc. A **62**, 416 (1949).
[2] for a review see M. Imada, A. Fujimori, and Y. Tokura, Rev. Mod. Phys. **70**, 1039 (1998).
[3] W. F. Brinkman and T. M. Rice, Phys. Rev. B **2**, 4302 (1970).
[4] for a review see A. Georges, G. Kotliar, W. Krauth, and M. J. Rozenberg, Rev. Mod. Phys. **68**, 13 (1996).
[5] H. Yokoyama, M. Ogata, and Y. Tanaka, J. Phys. Soc. Jpn. **75**, 114706 (2006).
[6] H. R. Krishnamurthy, C. Jayaprakash, S. Sarker, and W. Wenzel, Phys. Rev. Lett. **64**, 950 (1990).
[7] L. F. Tocchio, H. Feldner, F. Becca, R. Valenti, and C. Gros, Phys. Rev. B **87**, 035143 (2013).
[8] J. Kokalj and R. H. McKenzie, Phys. Rev. Lett. **110**, 206402 (2013).
[9] H. T. Dang, X. Y. Xu, K.-S. Chen, Z. Y. Meng, and S. Wessel, Phys. Rev. B **91**, 155101 (2015).
[10] Y. Shimizu, K. Miyagawa, K. Kanoda, M. Maesato, and G. Saito, Phys. Rev. Lett. **91**, 107001 (2003).
[11] for a review see B. J. Powell and R. H. McKenzie, Rep. Prog. Phys. **74**, 056501 (2011).
[12] J. Kokalj and R. H. McKenzie, Phys. Rev. B **91**, 205121 (2015).
[13] J. Kokalj and R. H. McKenzie, Phys. Rev. B **91**, 125143 (2015).
[14] M. Capone, M. Fabrizio, C. Castellani, and E. Tosatti, Rev. Mod. Phys. **81**, 943 (2009).
[15] A. Y. Ganin, Y. Takabayashi, P. Jeglič, D. Arčon, A. Potočnik, P. J. Baker, Y. Ohishi, M. T. McDonald, M. D. Tzirakis, A. McLennan, G. R. Darling, M. Takata, M. J. Rosseinsky, and K. Prassides, Nature **466**, 221 (2010).
[16] W. Kohn, Phys. Rev. **133**, A171 (1964).
[17] G. Kemeny and L. G. Caron, Rev. Mod. Phys. **40**, 790 (1968).
[18] T. A. Kaplan, P. Horsch, and P. Fulde, Phys. Rev. Lett. **49**, 889 (1982).
[19] T. Sato and H. Tsunetsugu, Phys. Rev. B **90**, 115114 (2014).
[20] P. Phillips, Rev. Mod. Phys. **82**, 1719 (2010).
[21] S. Zhou, Y. Wang, and Z. Wang, Phys. Rev. B **89**, 195119 (2014).
[22] Z. Lenarčič and P. Prelovšek, Phys. Rev. Lett. **108**, 196401 (2012).
[23] L. Laloux, A. Georges, and W. Krauth, Phys. Rev. B **50**, 3092 (1994).
[24] J. Bauer and A. C. Hewson, Phys. Rev. B **76**, 035118 (2007).
[25] F. Kagawa, T. Itou, K. Miyagawa, and K. Kanoda, Phys. Rev. Lett. **93**, 127001 (2004).
[26] P. Nozières and S. Schmitt-Rink, J. Low Temp. Phys. **59**, 195 (1985).
[27] M. Randeria, N. Trivedi, A. Moreo, and R. T. Scalettar, Phys. Rev. Lett. **69**, 2001 (1992).
[28] M. Keller, W. Metzner, and U. Schollwöck, Phys. Rev. Lett. **86**, 4612 (2001).
[29] M. Capone, C. Castellani, and M. Grilli, Phys. Rev. Lett. **88**, 126403 (2002).
[30] for a review see R. Micnas, J. Ranninger, and S. Robaszkiewicz, Rev. Mod. Phys. **62**, 113 (1990).
[31] K. Seki, R. Eder, and Y. Ohta, Phys. Rev. B **84**, 245106 (2011).
[32] J. E. Hirsch, Phys. Rev. B **31**, 4403 (1985).
[33] T. Schäfer, F. Geles, D. Rost, G. Rohringer, E. Arrigoni, K. Held, N. Blümer, M. Aichhorn, and A. Toschi, Phys. Rev. B **91**, 125109 (2015).
[34] R. Bulla, Phys. Rev. Lett. **83**, 136 (1999).
[35] W. F. Brinkman and T. M. Rice, Phys. Rev. B **2**, 1324 (1970).
[36] We note that deviations from the exactly semicircular DOS (applicable for Bethe lattice with $z \to \infty$) due to used finite $z$ do not apply in our case since first hopping in $z$ possible directions is treated explicitly and all the subsequent hoppings have $z - 1$ possibilities. This, e.g., corresponds to changing $z/(z-1) \to 1$ in Eq. 2.26 in Ref. 35 and leading to the exactly semicircular DOS.
[37] B. Bernu, C. Lhuillier, and L. Pierre, Phys. Rev. Lett. **69**, 2590 (1992).
[38] L. Capriotti, A. E. Trumper, and S. Sorella, Phys. Rev. Lett. **82**, 3899 (1999).
[39] T. Koretsune and M. Ogata, Phys. Rev. Lett. **89**, 116401 (2002).
[40] J. Kokalj and P. Prelovšek, Eur. Phys. J. B **63**, 431 (2008).
[41] S. Watanabe and M. Imada, J. Phys. Soc. Jap. **73**, 1251 (2004).
[42] A. N. Kocharian, N. Kioussis, and S. H. Park, J. Phys.: Condens. Matter **13**, 6759 (2001).
[43] J. Jaklič and P. Prelovšek, Adv. Phys. **49**, 1 (2000).
[44] Z. Weihong, R. H. McKenzie, and R. R. P. Singh, Phys. Rev. B **59**, 14367 (1999).
[45] T. Mizusaki and M. Imada, Phys. Rev. B **74**, 014421 (2006).
[46] T. Watanabe, H. Yokoyama, Y. Tanaka, and J. Inoue, Phys. Rev. B **77**, 214505 (2008).
[47] T. Yoshioka, A. Koga, and N. Kawakami, Phys. Rev. Lett. **103**, 036401 (2009).
[48] H.-Y. Yang, A. M. Läuchli, F. Mila, and K. P. Schmidt, Phys. Rev. Lett. **105**, 267204 (2010).
[49] J. Reuther and R. Thomale, Phys. Rev. B **83**, 024402 (2011).
[50] G. Li, A. E. Antipov, A. N. Rubtsov, S. Kirchner, and W. Hanke, Phys. Rev. B **89**, 161118 (2014).
[51] M. Laubach, R. Thomale, C. Platt, W. Hanke, and G. Li, Phys. Rev. B **91**, 245125 (2015).